\documentclass[twocolumn]{aastex62}

\usepackage{amsfonts}
\usepackage{amsmath}
\usepackage{amssymb}
\usepackage{bm}
\usepackage{color}
\usepackage{dcolumn}
\usepackage{epsfig}
\usepackage{eucal}
\usepackage{float}
\usepackage{graphicx}
\usepackage[latin1]{inputenc}
\usepackage{microtype}
\usepackage{subfigure}
\usepackage{tikz}
\usepackage{ulem}
\usepackage{verbatim}
\usepackage{xcolor}
\usepackage{cancel}

\begin{document}

\title{Future prospects on constraining neutrino cosmology with the Ali CMB Polarization Telescope}

\author{Dongdong Zhang}
\affiliation{Department of Astronomy, School of Physical Sciences, University of Science and Technology of China, Hefei, Anhui 230026, China}
\affiliation{CAS Key Laboratory for Research in Galaxies and Cosmology, University of Science and Technology of China, Hefei, Anhui 230026, China}
\affiliation{School of Astronomy and Space Science, University of Science and Technology of China, Hefei, Anhui 230026, China}

\author{Jiarui Li}
\affiliation{Department of Astronomy, School of Physical Sciences, University of Science and Technology of China, Hefei, Anhui 230026, China}
\affiliation{CAS Key Laboratory for Research in Galaxies and Cosmology, University of Science and Technology of China, Hefei, Anhui 230026, China}
\affiliation{School of Astronomy and Space Science, University of Science and Technology of China, Hefei, Anhui 230026, China}

\author{Jiaqi Yang}
\affiliation{Department of Astronomy, School of Physical Sciences, University of Science and Technology of China, Hefei, Anhui 230026, China}
\affiliation{CAS Key Laboratory for Research in Galaxies and Cosmology, University of Science and Technology of China, Hefei, Anhui 230026, China}
\affiliation{School of Astronomy and Space Science, University of Science and Technology of China, Hefei, Anhui 230026, China}

\author{Yufei Zhang}
\affiliation{Department of Astronomy, School of Physical Sciences, University of Science and Technology of China, Hefei, Anhui 230026, China}
\affiliation{CAS Key Laboratory for Research in Galaxies and Cosmology, University of Science and Technology of China, Hefei, Anhui 230026, China}
\affiliation{School of Astronomy and Space Science, University of Science and Technology of China, Hefei, Anhui 230026, China}

\author{Yi-Fu Cai}
\affiliation{Department of Astronomy, School of Physical Sciences, University of Science and Technology of China, Hefei, Anhui 230026, China}
\affiliation{CAS Key Laboratory for Research in Galaxies and Cosmology, University of Science and Technology of China, Hefei, Anhui 230026, China}
\affiliation{School of Astronomy and Space Science, University of Science and Technology of China, Hefei, Anhui 230026, China}

\author{Wenjuan Fang}
\affiliation{Department of Astronomy, School of Physical Sciences, University of Science and Technology of China, Hefei, Anhui 230026, China}
\affiliation{CAS Key Laboratory for Research in Galaxies and Cosmology, University of Science and Technology of China, Hefei, Anhui 230026, China}
\affiliation{School of Astronomy and Space Science, University of Science and Technology of China, Hefei, Anhui 230026, China}

\author{Chang Feng}
\affiliation{Department of Astronomy, School of Physical Sciences, University of Science and Technology of China, Hefei, Anhui 230026, China}
\affiliation{CAS Key Laboratory for Research in Galaxies and Cosmology, University of Science and Technology of China, Hefei, Anhui 230026, China}
\affiliation{School of Astronomy and Space Science, University of Science and Technology of China, Hefei, Anhui 230026, China}

\email{yifucai@ustc.edu.cn}
\email{wjfang@ustc.edu.cn}
\email{changfeng@ustc.edu.cn}

\begin{abstract}

We forecast the constraints on the parameters of neutrino physics with the constructions of Cosmic Microwave Background (CMB) temperature, E-mode polarization and lensing spectra for the ground-based Ali Cosmic Microwave Background Polarization Telescope (AliCPT). 
To implement the forecast calculations, we perform numerical simulations which show that AliCPT would yield the result $\sigma (N_{\mathrm{eff}})=0.42$ and $\sigma (M_{\nu})=0.18$ for the first year observation. 
Moreover, we investigate how the instrumental parameters, such as noise level, full width at half maxima (FWHM), and sky coverage can affect the constraints on these two parameters.
Our forecasting results find that a large aperture telescope with a large sky survey strategy would significant improve the current constraints.

\end{abstract}



\section{Introduction}\label{sec:introduction}

In the past several decades, CMB observations have provided us with a lot of important information for our universe \citep{2020A&A...641A...6P, 2021PhRvL.127o1301A}, and it will continue to play a crucial role in searching for new physics in the future \citep{2021arXiv210512554C}. 
The Planck mission \citep{2020A&A...641A...6P} ushered in the era of precise observations of the early universe. 
Due to the relaxation of the size of the equipment, the ground-base CMB experiments is committed to achieving lower instrumental noise of the CMB power spectra, such as the Background Imaging of Cosmic Extragalactic Polarization 2 (BICEP2 )/Keck Array \citep{2018arXiv180702199T}, the Cosmology Large Angular Scale Surveyor (CLASS) \citep{2020JLTP..199..289D}, POLARBEAR/Simons Array \citep{2016JLTP..184..805S}, the Atacama Cosmology Telescope (ACT) \citep{2017JCAP...06..031L} and the South Pole Telescope (SPT) \citep{2011PASP..123..568C}.

In the next decade, the construction of new generation of CMB experiments will
be greatly increased. 
For space telescope project, the LiteBIRD \citep{2014JLTP..176..733M, 2018JLTP..193.1048S} has been selected as the second Strategic Large-class missions by the Japan Aerospace Exploration Agency. 
On the ground, the Simons Observatory (SO) \citep{2019JCAP...02..056A}, the South Pole Observatory (SPO) and the ambitious Stage-IV network of ground-based
observatories (CMB-S4) \citep{2016arXiv161002743A, 2019arXiv190704473A, 2019BAAS...51g.209C} will be built on Atacama and Antarctica in the
southern hemisphere. 
In the northern hemisphere, AliCPT \citep{2016SCPMA..59g.178C, 2017arXiv170909053L, 2017arXiv171003047L}, established in Tibet, China, can complement the CMB data observed from the northern hemisphere.
Due to its good Precipitable Water Vapor (PWV) performance in the observing
season of Ali \citep{2017arXiv170909053L}, AliCPT has an excellent geographical advantage of observation at 95/150GHz in the northern hemisphere. 
Ideally, its observable area can reach to 70\% of the sky. 
As the AliCPT mission will be a long-term construction project, we expect it to contribute to many important scientific goals in near future.

As a frontier research content\citep{2020PhRvL.125z1105S, 2021PhRvL.126a1301M, 2021PhRvL.127d1101K, 2021PhRvL.126t1801B}, the nature of neutrinos is listed as one major scientific goal of AliCPT. In particular, it is well known that the dynamical nature of our universe is sensitive to two parameters
of neutrino physics \citep{2014NJPh...16f5002L}, i.e. the effective species of neutrinos $N_{\mathrm{eff}}$ and the sum of neutrino masses $M_v$. 
The high-precision measurements of $N_{\mathrm{eff}}$ shall clarify whether there exist any other light relics beyond the Standard Model (SM) of particle physics, such as axions in the early universe \citep{2013JHEP...12..058B, 2016PhRvL.117q1301B}. 
The precise upper bound of $M_v$ can tell us whether the mass of the neutrinos is in the normal hierarchy or the inverted hierarchy (see Section 2 for details), which has a significant impact on the research beyond SM.
In the present study, we use the aforementioned two parameters of neutrino physics as the constraining target to forecast the constraining power of AliCPT. Then, we discuss the impact of different follow-up construction scenarios on its ability to improve parameter constraints. 
We provide theoretical analysis and guidance for the subsequent construction of AliCPT.

Our article is organized as follows. 
In Section 2 we briefly review the neutrino cosmology. 
Section 3 includes the description of the methodology and some equipment parameters we adopted in the detailed calculations. 
We report our forecast results and discussions of the different instrumental improvement schemes in Section 4. 
We conclude with a discussion on future perspectives in Section 5.

\section{Neutrino Cosmology} 

In this section, we briefly review the observational effects of two important parameters in neutrino cosmology ($N_{\mathrm{eff}}$ and $M_v$) on the CMB power spectra.

\subsection{Effective Species of Neutrinos}

The parameter of effective species of neutrinos $N_{\mathrm{eff}}$ is related to the density of cosmic neutrinos. Any departure of the standard value, $N_{\mathrm{eff}}= 3.046$ in $\Lambda$CDM, can be a signal of new physics \citep{2013JHEP...12..058B, 2016PhRvL.117q1301B, 2015PhRvD..92e5033C}. Such a deviation could affect the CMB power spectrum, which would be examined by high precision observations. 

The main imprint of $N_{\mathrm{eff}}$ left on the CMB concentrated in the CMB damping tail.
As the photon and electron are not coupled extremely tight, the inhomogeneity inside the mean free path of photon would be lowered. 
That causes an exponential suppress of the high-$\ell$ CMB power spectrum. 
The non-standard neutrino density would change the expansion rate in the radiation dominated era. 
Accordingly, the modified expansion rate changes the mean free path of photon and so does the damping effect.

Another observational effect of the modified neutrino density is the change on the redshift of the matter-radiation equality $z_{eq}$ through the following relation \citep{2020PhRvD.101f3536A},
\begin{equation}
  1 + z_{eq} = \frac{\Omega_m h^2}{\Omega_{_{\gamma}} h^2} \frac{1}{1 +
  0.2271 N_{\mathrm{eff}}},
\end{equation}
where $\Omega_{\gamma} h^2 = 2.47 \times 10^{- 5}$ can be inferred accurately from current CMB temperature measurements. 
As $z_{eq}$ can be measured by the CMB temperature power spectrum with the ratio between the heights of the first and third peaks, $N_{\mathrm{eff}}$ can be degenerate with $\Omega_m$. 
This degeneracy can be broken by another independent probes such as SNIa. 


Changing the density of neutrinos can also vary its gravitational effect \citep{2004PhRvD..69h3002B}. 
After the decouple around $T = 1$ MeV, the neutrinos propagate freely and thus the propagating speed is faster than the sound speed of the cosmic plasma. 
Consequently, the gravity of neutrinos outside the overdensed region tends to pull them out and affects their oscillations. This turns out to yield a shift of the CMB acoustic peak.

\subsection{The Sum of Neutrino Masses}

Fundamental physics pays a lot of attention to the neutrino masses \citep{2021PhR...914....1F}. It is known that, the flavour oscillation experiments manifest the existence of massive neutrinos \citep{1998PhRvL..81.1562F, 2002PhRvL..89a1301A, 2005PhRvL..94h1801A, 2008PhRvL.101m1802A}, which has three different massive eigenstates, $m_1$, $m_2$ and $m_3$. Unfortunately, oscillation experiments are not able to judge which one is larger between $m_1$ and $m_3$. This flexibility about the mass ordering results in two scenarios: normal hierarchy ($\Delta m_{31}^2 > 0$) and inverted hierarchy ($\Delta m_{13}^2 > 0$) \citep{2016JCAP...11..035H}. For example, the measurement at a confidence level of $3\sigma$ from \citep{2012PhRvD..86g3012F} give the results as follows, 
\begin{align}
 &\Delta m_{21}^2 = 7.62_{-0.50}^{+0.58} \times 10^{-5} eV^2 ~, \\
 &\Delta m_{31}^2 = 2.55_{-0.24}^{+0.19} \Big( -2.43_{-0.22}^{+0.21} \Big) \times 10^{-3} eV^2 ~,
\end{align}
and the value in parenthesis is for the inverted hierarchy, in which we can estimate the minimum sum of neutrino masses is roughly 0.06 eV (normal hierarchy) or 0.1 eV (inverted hierarchy).

Although the measurements above are very precise, it is unattainable to measure the absolute mass value of each neutrino species, and consequently the total mass of neutrinos \citep{2014JHEP...03..028B}. It might be possible to directly measure the absolute value of the neutrino mass with certain techniques, such as the Beta decay experiment \citep{2019arXiv191004688G}, but so far the experimental accuracy is 
unable to distinguish the mass hierarchies. 
Moreover, some other cosmological observations can also measure the sum of neutrino masses as an independent approach \citep{2011ARNPS..61...69W, 2006PhR...429..307L}, which provides another possibility of breaking the degeneracy of the mass hierarchies. 

Neutrinos were relativistic in the early universe, which contributed to the energy density of radiation, but later became non-relativistic and turn into part of the energy density of matter. Therefore, massive neutrinos affect the expansion history of the universe and have an observable impact on CMB. On the other hand, massive neutrinos which propagate freely after the decouple with large thermal velocity would suppress the growth of large-scale structure. This physical effect leaves observable imprints on both the CMB angular power spectra and the matter power spectrum.

In this work, we would like to briefly review how neutrino masses impact the CMB power spectra in the following three aspects.
Firstly, the variation of the redshift of matter-to-radiation equality changes the amplitude and the location of the power spectrum peaks. 
Secondly, the Robertson-Walker radius of the last scattering surface $d_A (z_{dec}) $ can be effected by changing the energy density of matter at late times, which also determines the position of CMB power spectrum peaks. 
Thirdly, owing to the late Integrated Sachs-Wolfe effect, the gradient of the low-$\ell$ tail of the CMB spectrum can be affected by the non-relativistic matter density. However, the cosmic variance makes it hard to measure this effect.   

Compared with the matter power spectrum $P(k,z)$, the impact of the sum of neutrino masses on the CMB power spectra is weaker, which makes it more difficult to be observed. Nonetheless, constraining $M_\nu$ is still feasible by combining the data from CMB power spectra with other large-scale structure surveys, such as the galaxy clustering, weak lensing and CMB lensing.

\section{Methodology for AliCPT}  

With AliCPT's equipment parameters, we simulate the CMB TT, TE, EE and lensing power spectra. We use the best-fit parameters of Planck 2018\footnote{The values of cosmological parameters are given as follows:
$\omega_\mathrm{b}=0.02242$, $\omega_\mathrm{cdm}=0.11933$, $\theta_s=1.04101$, $\ln10^{10}A_s=3.047$, $n_s=0.9665$, $\tau_{\mathrm{reio}}=0.0561$.} \citep{2020A&A...641A...6P} and assume that the sum of neutrino masses is 0.06 eV with three degenerate massive neutrinos as our fiducial model based on the $\Lambda$CDM+$M_{\nu}$ model and the $\Lambda$CDM+$N_{\mathrm{eff}}$ model, respectively. The instrumental parameters are listed in Table 1. We use the instrumental parameters of the first observation season in our calculations, and assume a 8.6 $\mu$K-arcmin noise level which is estimated from the following formula \citep{2017arXiv171003047L},
\begin{equation}
  w^{- 1} = \frac{4 \pi f_{sky} NET^2}{t_{obs} N_{\det}} ~.
\end{equation}
In the above equation, the noise equivalent temperature (NET) is determined by the detector performance, instrument design and atmospheric condition. Moreover, $t_{obs}$ is the effective observation time, which can be estimated to be 1000 hours for each year. $N_{\det}$ is the number of detectors. Note that for polarization, NET should be multiplied by the coefficient of $\sqrt{2}$ since each polarization signal requires two orthogonal linear polarization detectors.

We employ the public MCMC code Montepython \citep{2013JCAP...02..001A} and the public Boltzmann code CLASS \citep{2011JCAP...07..034B} for calculations. We choose the Metropolis Hastings algorithm as our sampling method. For each model, we run MCMC calculation until R-1 for each fundamental parameter reach to the order of 0.001, according to the Gelman-Rubin criterion \citep{1992StaSc...7..457G}.

For the CMB temperature and polarization power spectra, we assume that the noise on T and E mode are
statistically independent, which indicates that $N^{TE}_l = 0$. The noise
spectra on TT and EE in the case of multiple channels can be described as
\citep{1995PhRvD..52.4307K,2016JCAP...03..052E}:
\begin{align}
  & N^{EE}_l = \left[ \sum_v w_{E, \nu} \exp \left( - l (l + 1)
  \frac{\theta_\mathrm{{FWHM, \nu}}^2}{8 \ln 2} \right) \right], \\
  & N^{TT}_l = \frac{1}{2} N^{EE}_l ~.
\end{align}
Taking into account the foreground contamination,  we assume $l_{\min} = 20$
follow the prescription in \citep{2016JCAP...03..052E}. For $l_{\max}$, we estimate it to be the reciprocal of the best FWHM in the two channels of AliCPT. In our calculation,
it is assumed that $l_{\max} = 1000$.

For the lensing reconstructed field, we use the quadratic estimation method \citep{2003PhRvD..67h3002O} to compute the error on the power spectrum. We use the publicly available code quicklens\footnote{https://github.com/dhanson/quicklens} to implement the calculations.
In addition, we combine AliCPT and Placnk to study the improvement on the parameter constraints. Since AliCPT has no real data, we use the mock spectra of Planck which were provided in \citep{2019JCAP...01..059B} to perform the forecasting.

\begin{table}[ht]
\centering
\caption{Instrumental parameters for AliCPT. Schedule for the installation of number of modules and detectors. }
\label{tab:default}
\begin{tabular}{c|c|c|c|c}
\hline
      Observation Years &  $\mathrm{NET}(\mu K \sqrt{s})$ & $N_{\mathrm{\rm{mod}}}$ & $N_{\mathrm{det}}$ & $f_{\mathrm{sky}}$ \\
\hline
     1 & 350 & 4 & 6,816 & $10\%$  \\
     2 & 350 & 4+4 & 13,632 & $10\%$ \\
     3 & 350 & 4+8 & 20,448 & $10\%$ \\
     4 & 350 & 4+12 & 27,264 & $10\%$ \\
\hline
\end{tabular}
\end{table}

\section{Results} 

By constructing the CMB TT, EE, TE power spectra and the lensing power spectrum, the constraining power of AliCPT on $N_{\mathrm{eff}}$ and $M_v$ are as follows: 
\begin{align}
  \sigma (N_{\mathrm{eff}}) & = 0.42 \quad ({\rm TT,~ TE,~ EE + lensing}) ~, \\
  \sigma (M_{\nu}) & = 0.18 \quad ({\rm TT,~ TE,~ EE + lensing}) ~.
\end{align}
In the previous calculations, we assumed that the $\ell$ range is 20\textless$\ell$\textless1000 for AliCPT. Considering that the actual construction of the instrument may be inconsistent with our assumptions during the real observation, we calculate the impact of changing the boundary conditions of $l_{min}$ and $l_{max}$ on neutrino cosmology. The calculation results are listed in the Table 2.

\begin{table} [ht]
	\begin{center}\footnotesize
		\begin{tabular}{|c|c|c|c|}
			\hline
			\multicolumn{4}{|c|}{$l_{min}=20$  } \\
			\hline
			$l_{max}$ & $800$  & $1000 $ & $1200$  \\
			\hline
			$\sigma (N_{\mathrm{eff}})$ & 0.51 & 0.42 & 0.42\\ 
			$\sigma (M_{\nu})$ & 0.21 & 0.18 & 0.18\\
			\hline
			\hline
			\multicolumn{4}{|c|}{$l_{max}=1000$  }    \\
			\hline
			$l_{min}$ & $20$  & $50 $ & $100$ \\
			\hline
			$\sigma (N_{\mathrm{eff}})$ & 0.42 & 0.43 & 0.44\\ 
			$\sigma (M_{\nu})$ & 0.18 & 0.28 & 0.33\\
			\hline
		\end{tabular}
	\end{center}
	\caption{The impact of changing $l_{min}$ and $l_{max}$ on the parameter constraints on neutrino cosmology.}
	\label{tab:ext_cmb_specs}
\end{table}

Increasing $l_{max}$ will not change the constraints significantly, which means that we use the hypothesis $l_{max}=1000$ to be reasonable. Increasing $l_{min}$ will reduce the constraining power of $M_{\nu}$ remarkably, while the effect on $N_{\mathrm{eff}}$ is marginal. This can be easily understood as discussed in Section 2, increasing $l_{min}$ will lose information from the late Integrated Sachs-Wolfe effect, thus weakening the constraint of $M_{\nu}$. However, this does not affect the constraint of $N_{\mathrm{eff}}$, because $N_{\mathrm{eff}}$ is more sensitive to the damping tail on the small scales of CMB.

Comparing AliCPT with the constraints by using Planck mock likelihood, $\sigma (N_{\mathrm{eff}})$=0.19 and $\sigma (M_{\nu})$=0.097 (TT,TE,EE + lensing), it is clear that the observation data of AliCPT in the first year is not outstanding in constraining neutrino physics. However, AliCPT provides CMB power spectra which have far lower noise level than Planck's, which can be used to enhance the joint constraining power. We use CMB TT, EE, TE power spectra of AliCPT to replace the range of $l= 20-1000$ on the corresponding 10\% sky coverage of Planck, while the noise of the combined lensing spectrum is constructed by the equation $\frac{1}{N^{dd}_l} = \frac{1}{N^{dd}_{1l}} + \frac{1}{N^{dd}_{2l}}$, in which $N^{dd}_{1l}$ \& $N^{dd}_{2l}$ means the lensing noise spectra for Planck and AliCPT, respectively. The results are as follows:
\begin{align}
  \sigma (N_{\mathrm{eff}}) & = 0.19 \quad ({\rm TT,~ TE,~ EE + lensing}) ~, \\
  \sigma (M_{\nu}) & = 0.088 \quad ({\rm TT,~ TE,~ EE + lensing}) ~.
\end{align}
After Planck combined with AliCPT, the optimization of constraint on $\sigma (N_{\mathrm{eff}})$ is not significant, but there is an improvement on $\sigma (M_{\nu})$ by about 9.3\%. The reason is the same as mentioned before, CMB is more sensitive to $N_{\mathrm{eff}}$ on small scales while more sensitive to $M_{\nu}$ on large scales.

Considering that currently AliCPT project is still under
construction, it is necessary to explore the impact of different subsequent construction plans on the scientific goals. We separately calculated the influence of different strategy
choices to improve device parameters on neutrino cosmological constraints.
Specifically, we can consider to improve the constraining accuracy by changing the noise
level, full width at half maxima (FWHM) and sky coverage $f_{sky}$.

In Table 3, we calculated the impact of changing different equipment parameters on the constraints of $N_{\mathrm{eff}}$ and $M_{\nu}$. The values of the first column of data are worse than the previous estimates of instrumental parameters, which are used as "conservative estimates" in actual observations.

\begin{table}[ht]
\caption{The influence of changing instrumental parameters on parameter constraints. The upper three rows show the constraining accuracy of $N_{\mathrm{eff}}$ and $M_{\nu}$ in the process of gradually reducing $w^{- 1/2}$ from 8.6*2 to 8.6/8 $\mu K$-arcmin under the $\Lambda$CDM+$M_{\nu}$ model and the $\Lambda$CDM+$N_{\mathrm{eff}}$ model, respectively. The middle three rows reflect the impact of changing FWHM, "FWHM multiplier" means the multiplier factor for the original parameter value (19 arcmin and 11 arcmin for tow channels). Meanwhile, $l_{max}$ gradually increased from 700 to 1000, 1200, 1500, 2500. The bottom three columns show the influence of elevate $f_{sky}$.}
\begin{center}
\setlength{\tabcolsep}{1.5mm}{
\begin{tabular}{c|c|c|c|c|c}
\hline
      $w^{- 1/2}$ ($\mu K$-arcmin) & 8.6*2 & 8.6 & 8.6/2 & 8.6/4 & 
      8.6/8 \\
\hline
     $\sigma (N_{\mathrm{eff}})$ & 0.48 & 0.42 & 0.38 & 0.37 & 0.36  \\
     $\sigma (M_{\nu})$ & 0.30 & 0.18 & 0.14 & 0.13 & 0.13  \\
\hline
\hline
      FWHM multiplier & 1.5 & 1.0 & 0.8 & 0.6 & 
      0.4  \\
\hline
     $\sigma (N_{\mathrm{eff}})$ & 0.58 & 0.42 & 0.30 & 0.27 & 0.14  \\
     $\sigma (M_v)$ & 0.27 & 0.18 & 0.16 & 0.14  & 0.11 \\
\hline
\hline
      $f_{sky}$ & 0.07 & 0.1 & 0.2 & 0.3 & 0.4 \\
\hline
     $\sigma (N_{\mathrm{eff}})$ & 0.50 & 0.42 & 0.30 & 0.24 & 0.21 \\
     $\sigma (M_{\nu})$ & 0.22 & 0.18 & 0.14 & 0.12 & 0.11 \\
\hline
\end{tabular}}
\end{center}
\end{table}

As shown in Table 3, the accuracy of $\sigma (N_{\mathrm{eff}})$ increased by 14\% during the process of reducing the noise level from 8.6 $\mu$K-arcmin to 8.6/8 $\mu$K-arcmin. At the same time, the accuracy of $\sigma (M_{\nu})$ can be greatly improved from 0.18 to 0.13, which is also our expectation for AliCPT after several years of increasing the number of detectors and accumulating sky survey data. 
We draw the comparison of the signal and instrumental noise on the CMB TT and EE spectra in Figure 1 for the the first year observation of AliCPT, as well as the signal and noise of $ C^{dd}_l$ under different noise levels in Figure 2 to interpret this difference.

\begin{figure} [t]
\begin{center}
{\includegraphics[angle=0,scale=0.33]{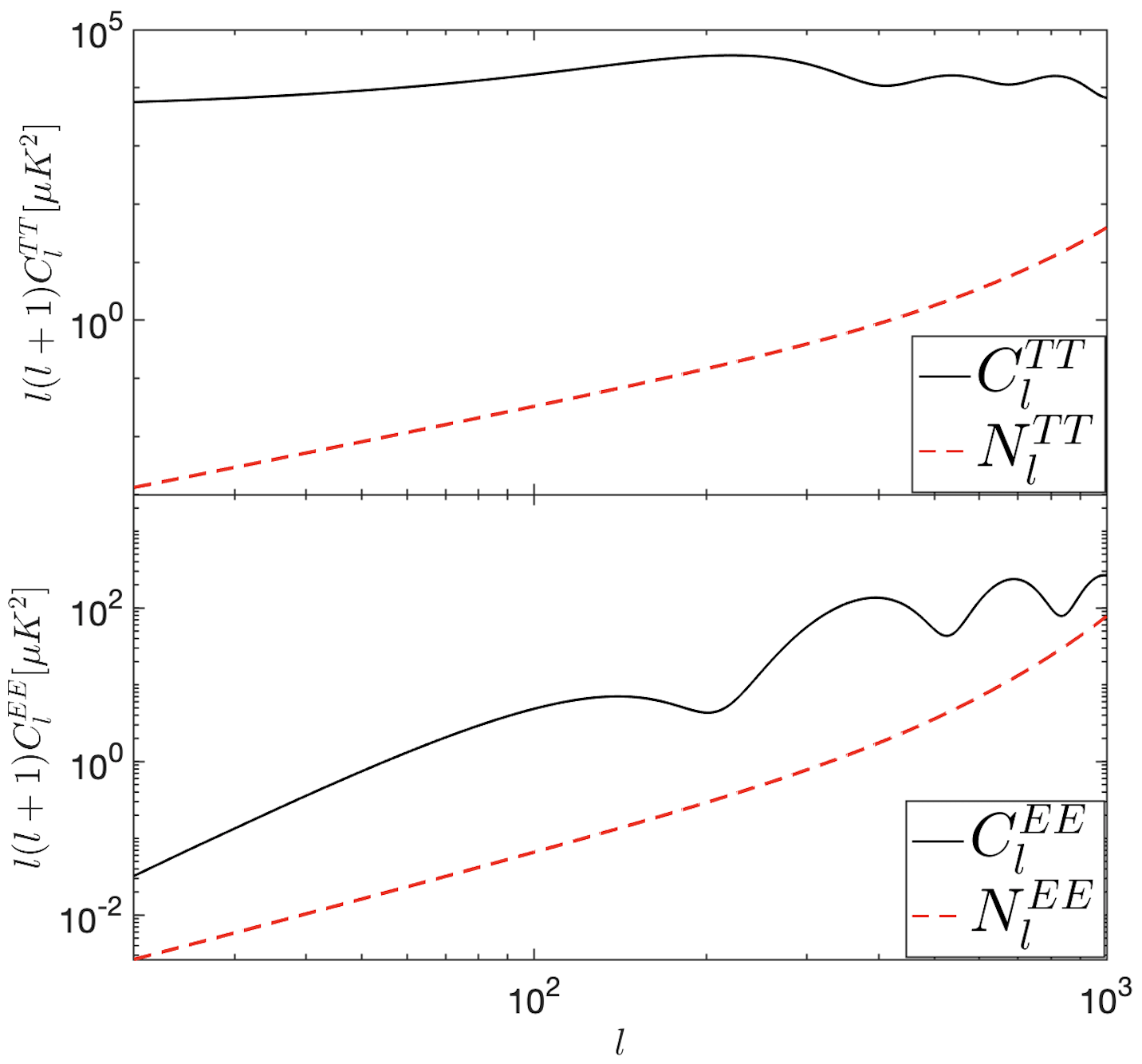}}
 \caption{\label{fig:analysis}  Signal and noise on the TT/EE spectra for the AliCPT. The black lines are signals of TT/EE spectra, and the red lines are instrumental noise spectra. }
\end{center}
\end{figure}

In Figure 1, it is easy to see that under the noise level of 8.6 $
\mu$K-arcmin, the error of the CMB TT spectrum in any $l$ region is dominated by cosmic variance. Therefore, it is understandable that simply reducing the noise level has little effect on the constraints of cosmological parameters such as $N_{\mathrm{eff}}$ which is more sensitive to CMB power spectra data alone rather than lensing information.

For the CMB lensing reconstructed field, we calculated the deflection spectrum $C^{dd}_l$ and the noise spectrum $N^{dd}_l$ at each $l$ when $w^{-1/2}$ was reduced to 1/2, 1/4 and 1/8 times the original value, respectively, and the results
are shown in Figure 2.

\begin{figure} [t]
\begin{center}
{\includegraphics[angle=0,scale=0.33]{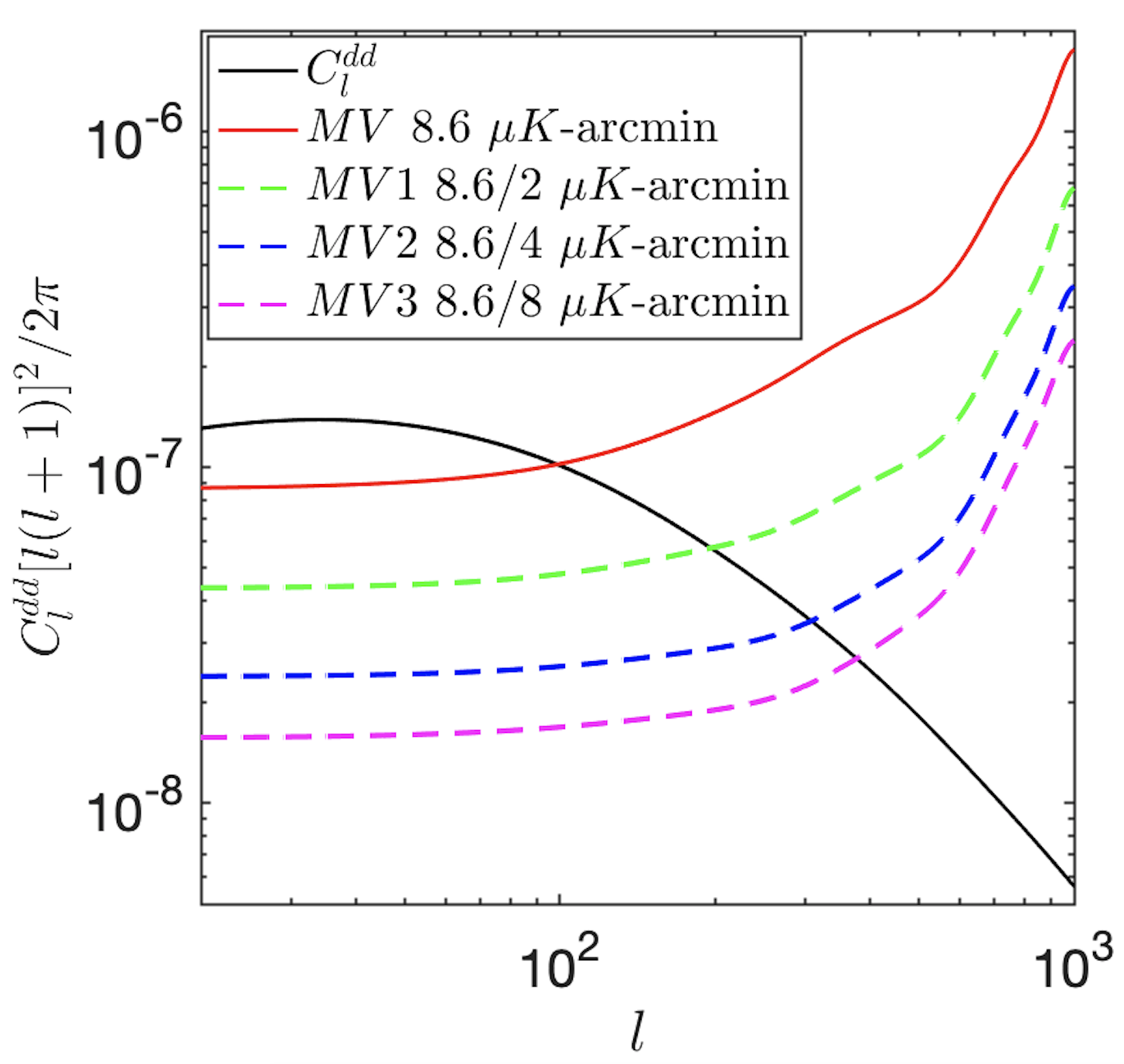}}
 \caption{\label{fig:analysis}  The impact of changing the noise level on $N^{dd}_l$ in the lensing deflection field spectrum. MV and MV1 to MV3 respectively represent lensing noise spectra $N^{dd}_l$ calculated based on $w^{-1/2}$ equal to 8.6, 8.6/2, 8.6/4 and 8.6/8 $\mu$K-arcmin, respectively.}
\end{center}
\end{figure}

It can be seen that $ N^{dd}_l$ still has a certain improvement in 8.6/2 and 8.6/4 $\mu$K-arcmin of the cases in Figure 2, but there is only slight improvement afterwards. Compared with the strength of the signal, the reduction of noise has an objective benefit. Since the constraint on $M_{\nu}$ is highly dependent on the information of the large-scale structure, this can explain why reducing noise level can significantly improve the constraint on $M_{\nu}$ compared the original value with the cases of 8.6/2 and 8.6/4 $\mu$K-arcmin, but the utility is no longer obvious afterwards.

On the other hand, Table 3 shows that changing FWHM dose have a significant improvement on the constraints, especially on $\sigma (N_{\mathrm{eff}})$. For $\sigma (N_{\mathrm{eff}})$, from 1 to 0.4 for the FWHM multipliers, the accuracy is improved from 0.42 to 0.14. Considering that the constraint on $\sigma (N_{\mathrm{eff}})$ of Planck mock likelihood is $0.19$, when FWHM is changed to $0.4$ times the original value $19$ arcmin and $11$ arcmin for two channels, we will be able to expect a better result of $\sigma (N_{\mathrm{eff}})$. Considering that we did not include the improvement of noise level and $f_{sky}$, this is a conservative conclusion.

For $\sigma (M_{\nu})$, from 1 to 0.4 for the FWHM multipliers, the
accuracy is improved from 0.18 to 0.11. Clearly it is not an
ideal choice to simply use CMB spectra and CMB lensing data to constraining $M_{\nu}$. 
The potential of the cross-correlation of CMB and large scale structure to constrain cosmological parameters have been demonstrated in literature,see e.g. \citep{2021PhRvD.103d3539Z}.
Since the sky survey region of AliCPT and DESI have a high degree of coincidence, we would expect it to be more significant to constraining $M_{\nu}$ if we use the cross-correlation combine AliCPT and DESI. But this part of the analysis is beyond the scope of our discussion.

From the above discussion on reducing the noise level and FWHM, we can see that even if the FWHM is only improved to 0.8 times the original value, a constraint of $N_{\mathrm{eff}}$ which is better than the best possible result of the former can be obtained. For the constraint of $M_{\nu}$, the effectiveness of the two methods is roughly the same. Considering that increasing resolution has significantly improvement of the constraints on all cosmological parameters, a smaller value of full width at half maxima for AliCPT is in principle expected.

Changing the sky coverage would have a significant improvement on parameter constraining power for a direct reason that it would reduce the cosmic variance immediately, which would also promise to be beneficial to other scientific goals like the detection of primordial gravitational waves \citep{2021JCAP...08..033L}. A discussion of the impact of changing sky coverage on constraining $N_{\mathrm{eff}}$ can be found in Section 4.1.1 of \cite{2019JCAP...02..056A}. 

The current plan of AliCPT is to conduct surveys in less than $14\%$ of the sky in order to maximize the use of detectors to reduce noise level and avoid invalid sky surveys on the Milky Way. If we consider increasing the scope of the sky survey, we would like to propose to build a CMB survey array in Ali, dubbed as AliCPT-Array, to conduct simultaneous sky surveys on multiple regions that are suitable for observation and expected to reach $35\%$ of sky coverage. Because of the universality of reducing the cosmological variance to improve the constraining power, it can also strengthen AliCPT's research capabilities for various scientific goals. We will do more work in follow-up research to analysis it.

\section{Conclusions} 

In this work we have applied the MCMC method to forecast the constraining ability of AliCPT on two main parameters of neutrino cosmology $N_{\mathrm{eff}}$ and $M_{\nu}$. We have performed calculations with the goal of improving the ability of the equipment to constrain these two parameters by reducing noise level, improving the resolution and sky coverage.
One conclusion based on our analysis is that, AliCPT alone cannot show superiority in the constraints of neutrino physics when compared with the Planck data. However, the combination of AliCPT and Planck can reduce $\sigma (M_{\nu})$ by 9.3\% even for the first year observation of AliCPT.

Moreover, we calculated the effects of different options for improving equipment parameters on AliCPT. We found that providing more small-scale information or increasing $f_{sky}$ on AliCPT is much better than only reducing noise level. The former is mainly because $N_{\mathrm{eff}}$ is very sensitive to the small-scale information of CMB. The latter is because increasing $f_{sky}$ can significantly reduce cosmic variance, which dominates the main source of CMB error as shown in Fig.~1, and it will help to achieve other scientific goals such as detecting primordial gravitational waves. We will perform more detailed research on this aspect in the follow-up study.

In summary, the high-precision CMB data provided by AliCPT will contribute to improve Planck's ability in constraining the neutrino cosmology even for the first year observation, especially for $M_{\nu}$. Our analysis show that, for the improvement of AliCPT, a smaller value of FWHM is in principle expected. On the basis of the existing equipment construction, we also expect that in the future a possible promotion of the AliCPT Array to survey multiple observable low noise areas in the same time during the annual observation season could significantly improve the related scientific output capacity.

\section*{Acknowledgements} 
We are grateful to Jacques Delabrouille, Aoxiang Jiang, Hong Li, Siyu Li, Yang Liu, Wentao Luo, Ao Wang, Qingqing Wang, Deliang Wu, and Pierre Zhang for valuable communications. 
This work is supported in part by the National Key R\&D Program of China (2021YFC2203100), by the NSFC (Nos. 11653002, 11961131007, 11722327, 11421303, 11773024, 12173036), by the National Youth Talents Program of China, by the Fundamental Research Funds for Central Universities, by the CSC Innovation Talent Funds, by the CAS project for Young Scientists in Basic Research (YSBR-006), by the CAS Interdisciplinary Innovation Team, by the China Manned Space Project with No. CMS-CSST-2021-B01, and by the USTC Fellowship for International Cooperation. 
We acknowledge the use of computing facilities of astronomy department, as well as the clusters {\it LINDA} and {\it JUDY} of the particle cosmology group at USTC.


\bibliography{main}
\end{document}